\documentclass[aps,pra,amsmath,amssymb,twocolumn,showpacs]{revtex4}

\usepackage{bm}
\usepackage{dcolumn}
\usepackage{graphicx}
\usepackage{lineno}
\usepackage{color}

\newcommand{\bdm}{\begin{displaymath}}
\newcommand{\edm}{\end{displaymath}}

\newcommand{\be}{\begin{equation}}
\newcommand{\ee}{\end{equation}}
\newcommand{\bea}{\begin{eqnarray}}
\newcommand{\eea}{\end{eqnarray}}

\newcommand{\mb}{\begin{pmatrix}}
\newcommand{\me}{\end{pmatrix}}

\begin{document}
\title{Nonlinear Spatial Focusing in Random Layered Media by Spectral Pulse Shaping}
\author{Alex C. Han and Valery Milner}
\affiliation{
Department of Physics and Astronomy\\
The University of British Columbia, Vancouver, Canada}

\begin{abstract}
We demonstrate numerically a method of focusing two-photon field inside one-dimensional random media. The approach is based on coherent control of backscattering achieved by adaptive spectral pulse shaping. The spectral phases of a femtosecond laser pulse are adjusted for the constructive interference of its backward-traveling components, resulting in an enhanced reflection from within the random system. A delayed forward-propagating second pulse overlaps with the controlled reflection, increasing the inter-pulse multi-photon field at a location determined by the delay between the two pulses. The technique is shown to be robust against the variations of the disorder, and to work with realistic pulse shaping parameters, hence enabling applications in controlling random lasing and multi-photon imaging in scattering materials.
\end{abstract}

\maketitle

\section{Introduction}

Focusing of light in random media is of interest from both the fundamental and applied perspectives. One important application of light focusing in disordered materials relates to the controllability of coherent light amplification known as random lasing \cite{Letokhov1968, Cao2003, Wiersma2008}. Lasing action in the presence of random light scattering has been theoretically analyzed and experimentally studied in multiple materials and spatial geometries \cite{Genack1994, Lawandy1994, Sha1994, Noginov1995, Wiersma1995, Cao1999, Frolov1999, Jiang2000}. The very nature of random lasing implies uncertainty in the characteristics of the lasing modes \cite{Andreasen2011}. The ability to control the lasing frequency of a random laser has therefore been actively investigated and accomplished, for example, by changing the temperature of the scattering medium \cite{Wiersma2001}, its structure \cite{Wiersma2008, Gottardo2008, Bardoux2011} and absorption properties \cite{ElDardiry2011}, the wavelength \cite{Shojaie2014} and the location of local pumping \cite{Cao1999, Wu2007, Bachelard2012, Hisch2013}. The latter approach is particularly appealing in the regime of strong scattering, when the emission modes are localized in space and can be selectively excited by choosing the location of the pump which best overlaps the target lasing mode \cite{Vanneste2001, Sebbah2002}.

Local pumping requires optical access to different parts of the gain material. This can be achieved by applying an external pump laser to different areas on the surface of a three-dimensional sample \cite{Cao1999}. However, for systems of lower dimensionality (especially one-dimensional (1D) random lasers \cite{Burin2002, Milner2005}, where the regime of light localization is most easily established), local pumping implies the ability to focus the pump light \textit{inside} the scattering medium. It has recently been implemented by pumping a 1D random structure from the side, i.e. perpendicularly to the direction of the disorder \cite{Bachelard2012, Bachelard2014}. Such pumping geometry, however, may not always be available, either because of the lack of optical access or due to the microscopic size of the sample.

Here, we propose an alternative approach, in which local pumping of a random layered medium is accomplished by means of spectral shaping of femtosecond pump pulses. The concept is similar to the spatio-temporal focusing of ultrashort laser pulses in three-dimensional scattering media \cite{Aulbach2011, Katz2011, Mosk2012}. There, controlling the spatial degrees of freedom of the incident wave has proven successful in simultaneous focusing of multiply-scattered light both in space and time. However, similarly to the pumping from the side of a sample, spatial shaping may not always be available for 1D structures with an optical access to their front layer only. In this situation, spectral shaping can be used as a powerful tool for controlling light propagation through, and interaction with, scattering materials \cite{VanBeijnum2011}. The success of using the spectral degree of freedom has been recently demonstrated in the linear regime of spatio-temporal focusing \cite{McCabe2011}. Optical nonlinearity, such as two-photon absorption (TPA), offers higher degree of controllability in random media due to the process of multi-photon intra-pulse interference \cite{Meshulach1998, Delacruz2004}.

\begin{figure}[t]
\includegraphics[scale=0.61]{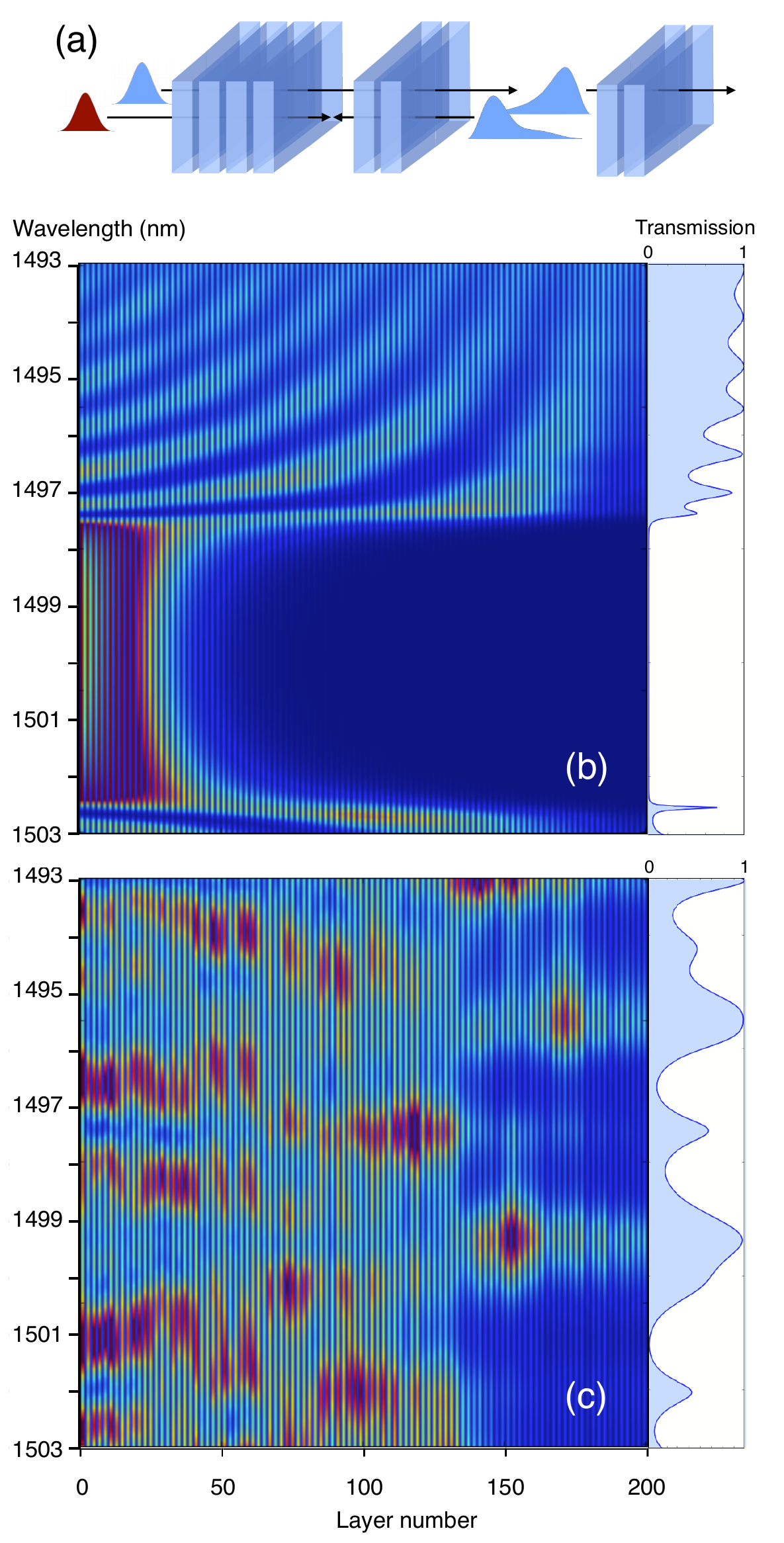}
\caption{(color online)
(\textbf{a}) Illustration of the proposed strategy for multi-photon focusing inside a one-dimensional ``random stack'', consisting of alternating layers of random thickness (dielectric blue slabs and air gaps). Spectral shaping of the control pulse (blue) is used to initiate strong localized backscattering from within the stack. The focusing of a two-photon field occurs at the location, where the backscattered control pulse meets a forward-propagating pump (red, drawn off-axis from the control pulse for clarity) (\textbf{b,c}) Field distribution inside an almost periodic stack (\textbf{b}, $\epsilon=0.1\%$) and a random stack (\textbf{c}, $\epsilon=20\%$) at different wavelengths. Band-edge and localized modes are visible inside a periodic and disordered sample, respectively, and are reflected by the peaks in the corresponding transmission spectra shown in the right panels.}
\label{mode_scan}
\end{figure}

The one-dimensional photonic structure of our choice is illustrated in Fig. \ref{mode_scan}\textbf{a}. Known as a ``random stack'', it consists of a series of layers with alternating refractive indices $n$: here, 100 dielectric layers with $n=1.1$ separated by air gaps ($n=1.0$). The layers are of uniform but random thickness, distributed in the range $[L(1-\epsilon), L(1+\epsilon)]$, where $L$ is 10 microns for dielectric layers and 1 micron for air layers, and $\epsilon $ is the dimensionless disorder parameter. Random layer thickness results in the complete phase scrambling of multiply scattered waves, whereas their interference in one dimension brings about the well known effect of light localization \cite{Berry1997}. Field intensity distributions inside two stacks, one with low and another with high degree of disorder are shown in Fig. \ref{mode_scan}~\textbf{b} and \textbf{c}, respectively, together with the corresponding transmission spectra (right panels).

The case of $\epsilon=0.1\%$ represents the photonic crystal limit with a stop band around 1500~nm. Localized edge modes can be seen around the boundaries between the pass and stop bands, but most of them are not sufficiently localized to be used in the proposed technique.
When the degree of disorder is increased to $\epsilon=20\%$, one can see the appearance of strongly localized modes associated with the randomly distributed transmission peaks. The parameters of the stack are chosen so as to result in the average mode line width of about 1~nm, about ten times larger than the resolution of a typical spectral pulse shaper \cite{Weiner2000}. Together with the ability to cover a number of localized modes under the spectrum of a typical femtosecond pulse, individual control of the spectral phases within each excited mode is essential for the proposed focusing scheme described below.

By inspecting the intensity distribution inside a random stack (Fig. \ref{mode_scan}\textbf{c}), one notes an obvious way of focusing light at different locations within the layered structure by means of narrowing the spectrum of a broadband excitation pulse and making it resonant with a single localized mode. However, such \textit{amplitude shaping} implies much lower peak intensities and correspondingly higher average power required to cross the lasing threshold \cite{Drane2015}.

An alternative \textit{phase-only shaping} is preferable from the standpoint of making use of the full bandwidth of an ultrashort pump pulse. Yet in the linear absorption regime, changing the spectral phases of the pulse amounts to altering its temporal profile but does not affect the total absorbed power \cite{Meshulach1999}. We therefore turn to a nonlinear optical process, such as two-photon absorption, as the enabling mechanism for the controlled focused excitation. Similarly to the well known colliding-pulse mode-locking scheme \cite{vanderZiel1981}, we propose to create a local TPA centre inside a random stack by overlapping two counter-propagating pulses: a backscattered ``control pulse'' and a forward-propagating ``pump pulse''. The two pulses meet at the location easily controlled by the time delay between them. Backscattering is necessary due to the assumed restricted optical access to the random sample, limited to its front layer only (left most layer in Fig.\ref{mode_scan}\textbf{a}). As we demonstrate below, spectral shaping is capable of creating the required strong backscattered pulse originating deep inside a disordered one-dimensional structure by controlling the phases of its localized modes. Furthermore, we show that the latter phases can be determined by means of an adaptive search algorithm which uses the energy of the pulse reflected from the front layer as the only feedback parameter.

\section{Two-photon Field Focusing}

The proposed technique achieves spatial focusing of the two-photon field in two steps. First, a control pulse, incident on the random stack along its axial direction, is spectrally shaped so as to backscatter from around the centre of the random sample. Then, a time-delayed pump pulse is injected along the same path. The central frequencies of the two pulses, $\omega _{c}$ and $\omega _{p}$, are assumed to be separated by more than the pulse bandwidths, as shown by their combined spectrum in the wavelength domain in Fig. \ref{spec}\textbf{a}. When the two pulses overlap in space and time, their two-photon field will acquire an inter-pulse sum-frequency component situated between the two individual second harmonics (shaded band in Fig. \ref{spec}\textbf{b}). The spatial focusing of this sum-frequency component within the random stack is determined by the delay between the two pulses, and can be used for the two-photon pumping of a gain medium, provided the latter absorbs light at frequency $\omega _{c}+\omega _{p}$.

\subsection{Controlled Backscattering}

At any position $z$ inside the stack, the field amplitude of the control pulse at frequency $\omega $ is a sum of the forward and backward-propagating components:
$
E(\omega,z) = a_j(\omega) e^{ik_jz} + b_j(\omega) e^{-ik_jz},
$
where $k_j = \omega n_j/c$ and $j$ is the layer number.
For a transform-limited control pulse incident on the stack, all spectral amplitudes $a_0(\omega)$ are in-phase, but multiple random scattering results in a significant phase scrambling between $a_j(\omega)$ [and hence, $b_j(\omega)$] amplitudes at locations deep inside the stack, causing dispersed reflection. However, if one constructs a spectral phase mask to pre-shape $a_0(\omega)$ before the control pulse enters the stack, such that the backward components acquire identical phases at a target location, a localized wave packet in the backward propagation direction can be generated.

To demonstrate the level of control over the backscattered pulse, we investigate one \textit{feedforward}-based method and two \textit{feedback}-based methods of finding the required pre-shaping mask. In the feedforward approach, the complex spectral amplitudes at the location of each layer are first calculated using the transfer matrix method \cite{Pendry1994} and \textit{a priori} knowledge about the random stack parameters. The required phase mask is the one that flattens the phases of the backscattered amplitudes $b_j(\omega)e^{-ik_jz}$ at a particular location $z$ (e.g. midpoint of layer $j$). The ease of finding such a mask stems from the fact that all pairs of $a_j$ and $b_j$ amplitudes are linearly dependent on the incident amplitude $a_0$, at any given frequency 
\footnote{
The field amplitudes at the entrance and exit sides are related via
$
( a_0 \quad b_0 )^\text{T} = M ( a_N \quad b_N )^\text{T}
$,
where $M = (T_{N-1}\dots T_1T_0)^{-1}$ involves the transfer matrices $T_j$ connecting layer $j$ and $j+1$ by 
$
( a_{j+1} \quad b_{j+1} )^\text{T} = T_j ( a_j \quad b_j )^\text{T}
$.
If all fields are incident to the front layer ($j=1$), then $b_N=0$, and it gives us
$
a_N =  a_0/M_{11},\quad
b_0 = M_{21} a_N = (M_{21}/M_{11}) a_0.
$
}.

In practice, however, the the geometry of the stack structure, i.e. the exact layer thicknesses and refractive indices, is usually unknown. In this case, the pre-shaping mask can be found by means of an adaptive feedback loop optimizing the backscattered field intensity
$
|\int_\omega b_j(\omega)e^{-i\omega (t-t_0)}e^{-ik_jz}d\omega|^2
$
either inside or outside the disordered sample. The internal field optimization may, again, be impractical. It will be analyzed below to serve as a reference for evaluating the performance of the second method based on optimizing the backward propagating field at the surface of the very front layer at a given time delay after the entrance of the control pulse. In what follows, we demonstrate how the three shaping methods produce the desired focusing of the two-photon field and analyze its dependence on several key physical quantities: the degree of disorder, the location of the focus within the stack, and the resolution of the pulse shaper.

\begin{figure}[t]
\includegraphics[scale=0.6]{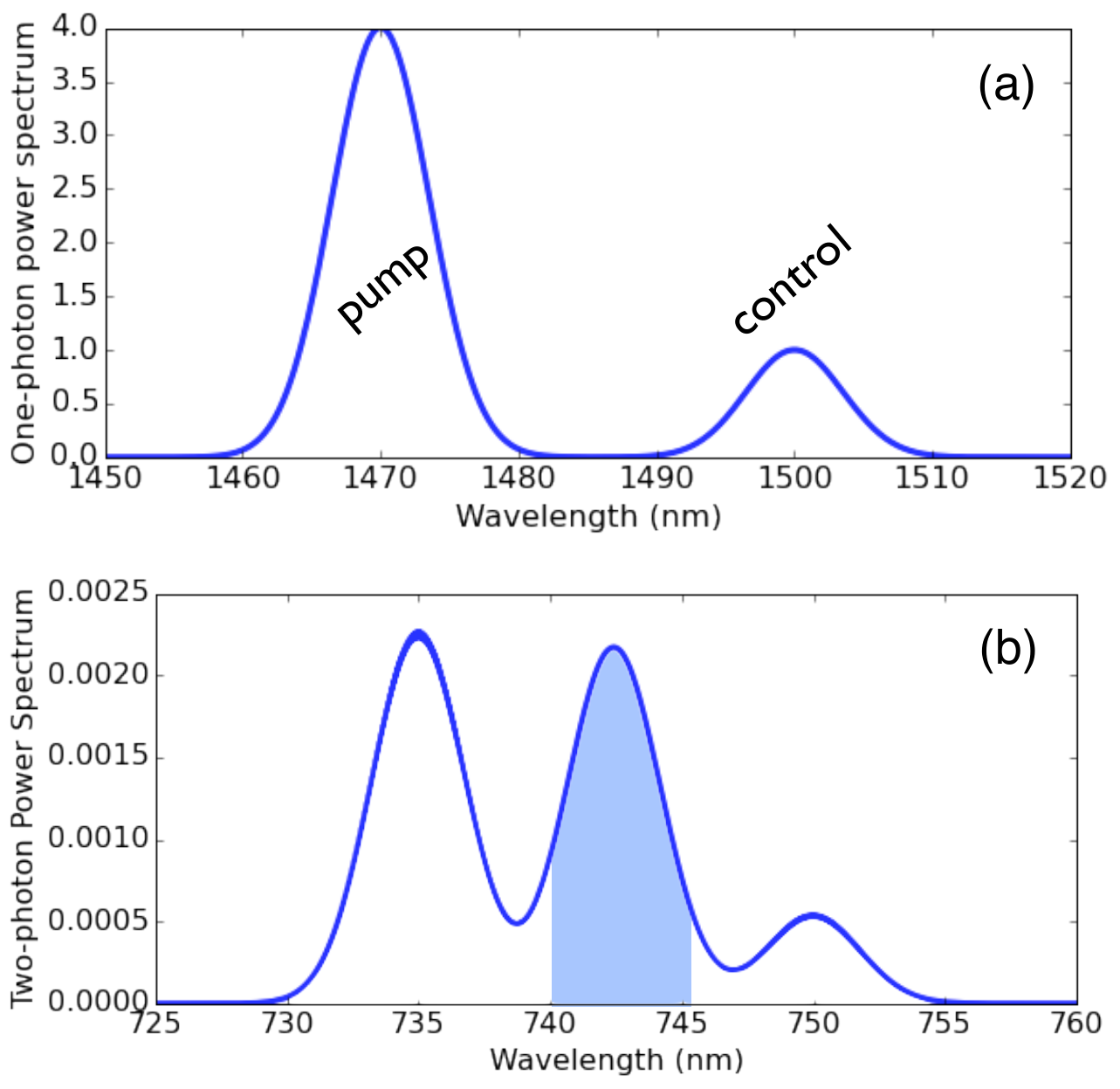}
\caption{One-photon (\textbf{a}) and two-photon (\textbf{b}) power spectra of control (centered at 1500nm) and pump (centered at 1480nm) pulses. The shaded area indicates the frequency range of the two-photon field used in the proposed focusing technique. It is calculated for the case of the transform-limited control and pump pulses overlapping in space and time.
}
\label{spec}
\end{figure}

Note that the central frequency of the control field was deliberately chosen in the middle of the stop band (cf. Fig. \ref{mode_scan} and \ref{spec}) to analyze our focusing technique in the periodic limit. Pump and control pulses had the same bandwidth, with pump being 4 times more intense in order to compensate for the weakening amplitude of the backscattered control light. The initial resolution of the pulse shaper was set at 0.07~nm.

\begin{figure*}[t]
\includegraphics[scale=0.44]{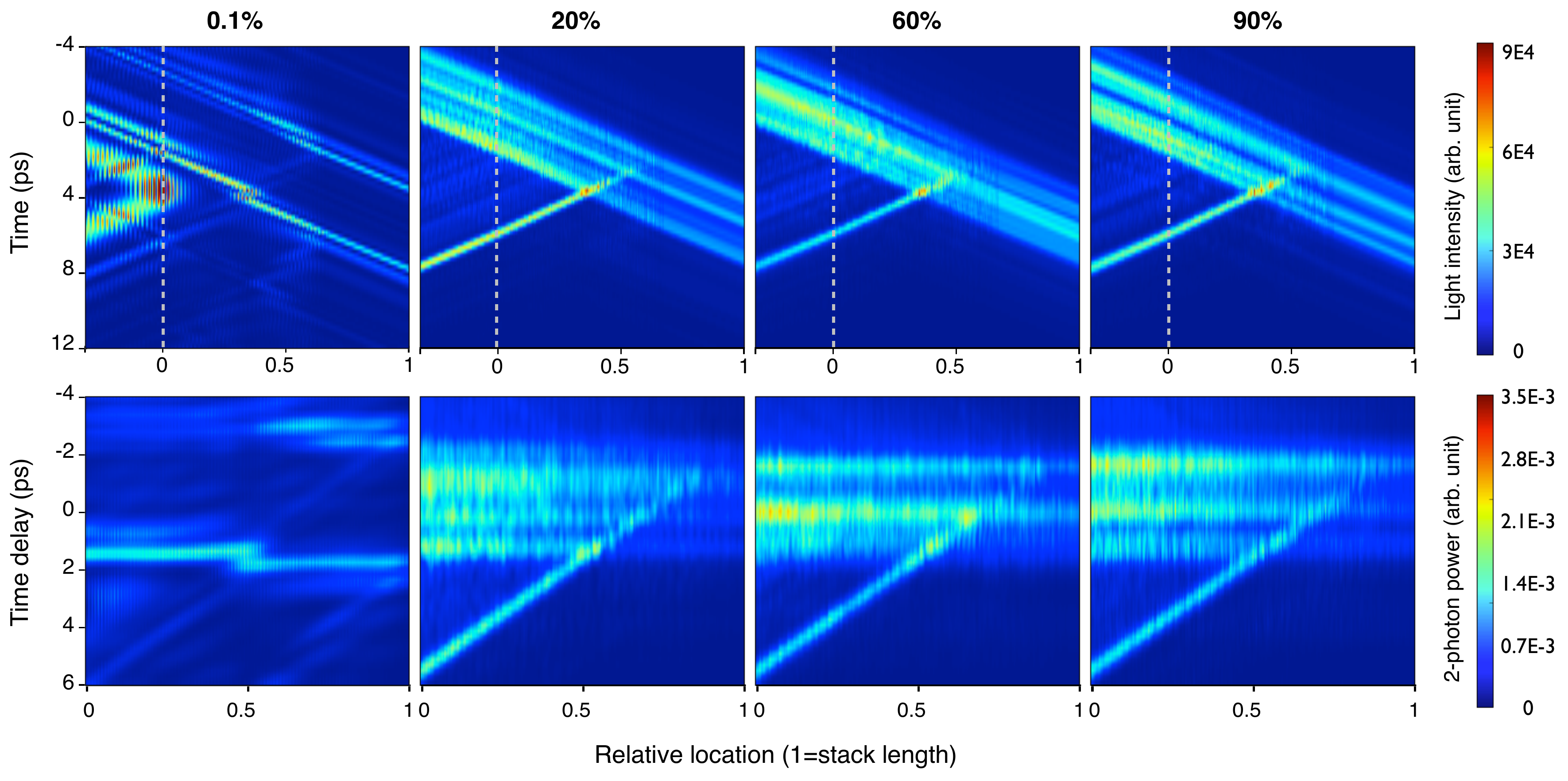}
\caption{(color online) Results of the \textit{feedforward} shaping method. One-photon (upper row) and two-photon (lower row) intensity distributions within random stacks with different degrees of disorder, $\epsilon$=0.1\%, 20\%, 60\%, 90\% (vertical columns). The one-photon field is for the spectrally shaped control pulse only. Vertical dashed lines indicate the front of the stack in the one-photon plots.
The two-photon field is plotted as a function of the delay time between the arrivals of the pump pulse and the unshaped control pulse. Each plot is an average of 10 realizations of disorder, i.e. layer thicknesses.}
\label{ff}
\end{figure*}

\subsection{Analysis of the Focusing Results}

\begin{figure*}
\includegraphics[scale=0.44]{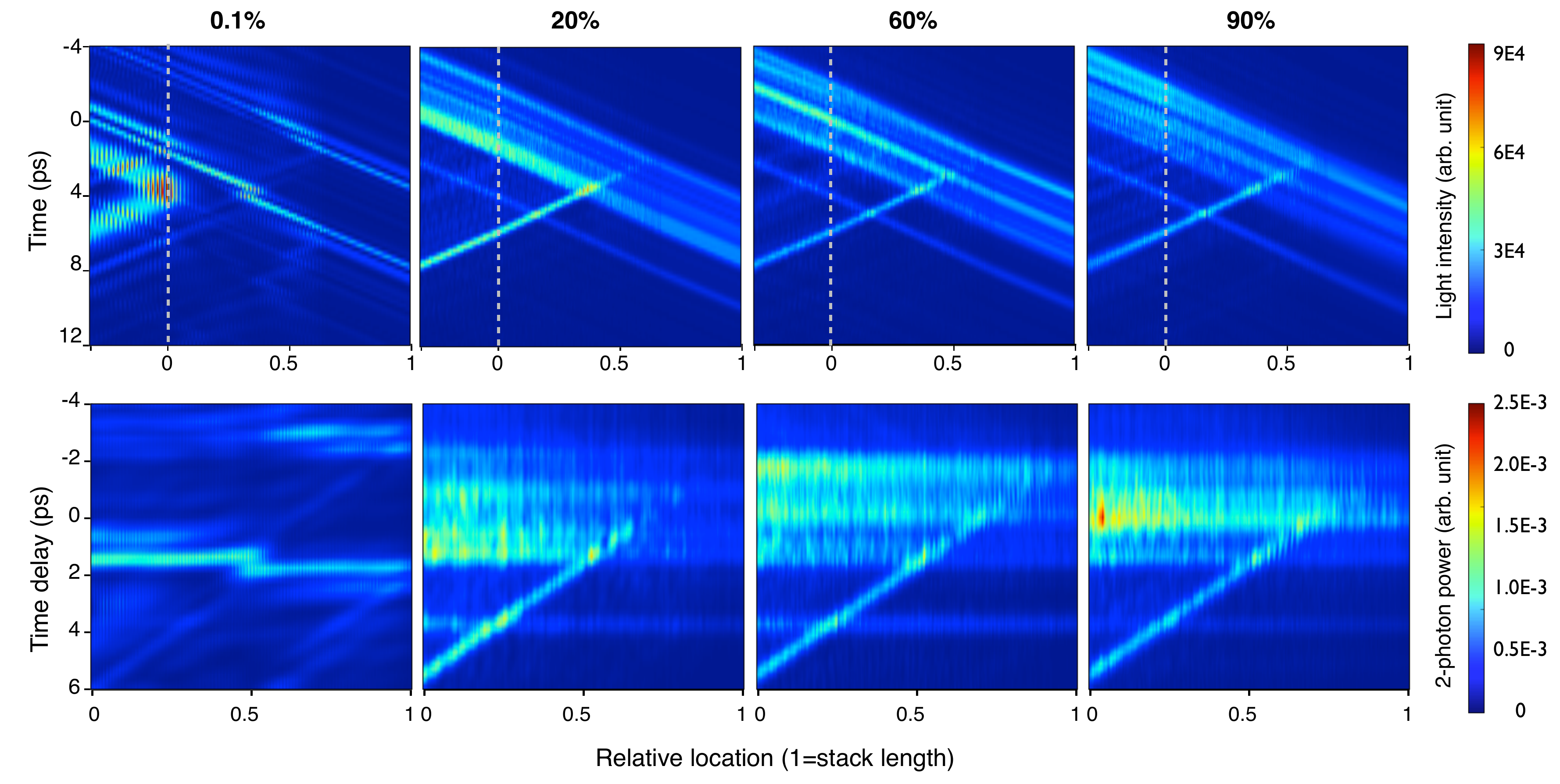}
\caption{(color online) Same as Fig. \ref{ff}, but for the \textit{feedback}-based shaping method, with the feedback parameter being the backscattered intensity at the 50th air layer. }
\label{fb100}
\includegraphics[scale=0.44]{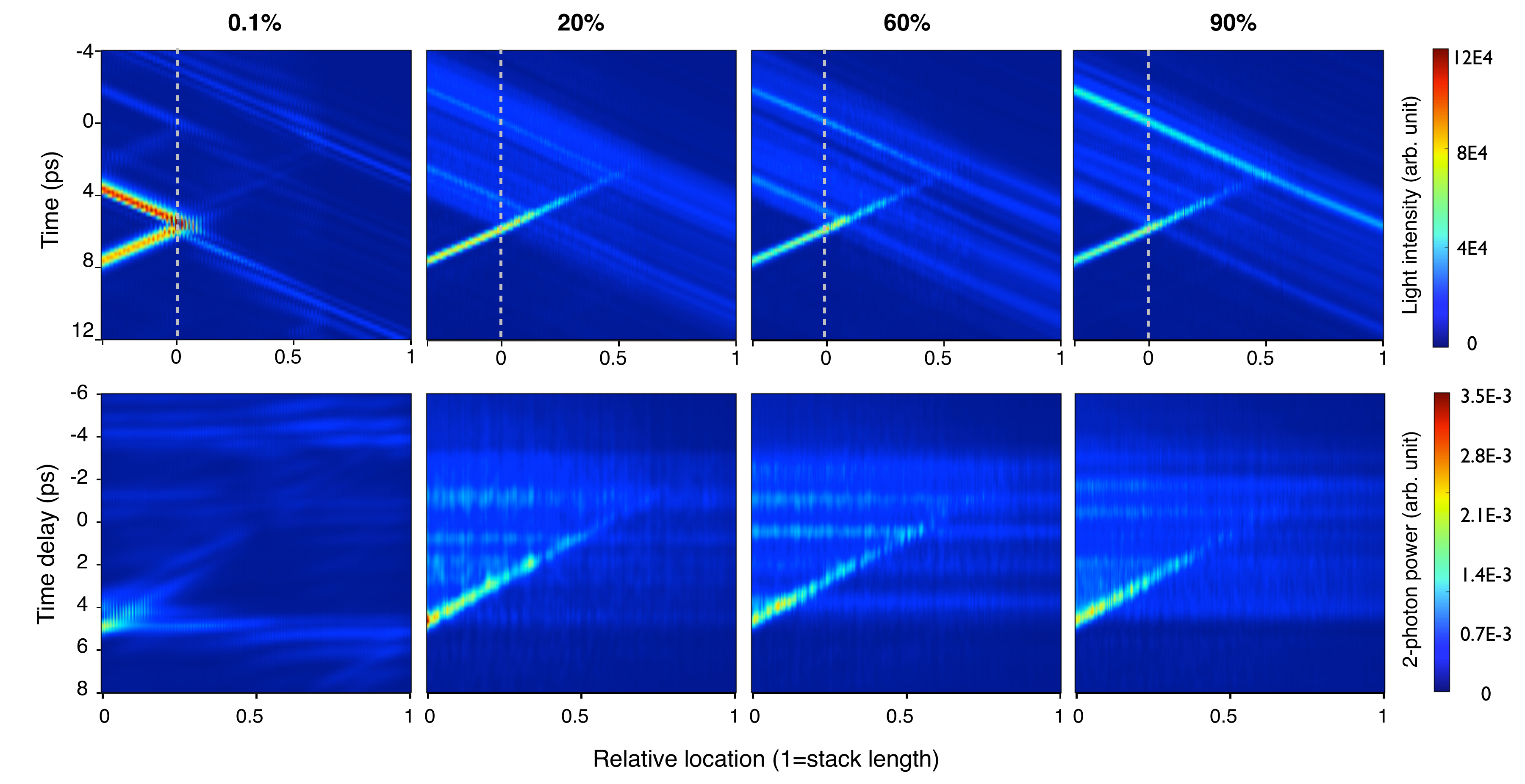}
\caption{(color online) Same as Fig. \ref{ff}, but for the \textit{feedback}-based shaping method, with the feedback parameter being the backscattered intensity at the front layer at $t=6$~ps ($t=0$ is the time of arrival of the unshaped control pulse).
}
\label{fb0}
\end{figure*}

\subsubsection{Dependence on the degree of disorder}

The performance of the three focusing methods discussed above is illustrated in Figures \ref{ff}, \ref{fb100} and \ref{fb0}. In the upper row we plot the trajectory of a control pulse, i.e. the one-photon field intensity distribution as a function of the propagation time. Time zero corresponds to the arrival time of an \textit{unshaped} control pulse at the front layer. The bottom row of plots shows the two-photon intensities for the shaded frequency band of Fig. \ref{spec}\textbf{b}, as a function of the control-pump delay time. Pairs of such plots (different columns) are then compared across different degrees of randomness expressed as a relative variation of thickness ($\epsilon $) of both the dielectric layers and the air gaps. Each plot is averaged over 10 realizations of the disorder.

In the feedforward shaping method, the phases of the backward propagating components are made equal at the midpoint of the 50th air layer. As can be seen in the upper row of Fig. \ref{ff}, the pre-shaping is reflected by the smearing of the control pulse in time prior to its entrance to the stack. The successful generation of a backscattered pulse from the middle of the stack is clearly visible across moderate to large layer thickness variations ($\epsilon$=20\%, 60\%, 90\%). The applied shaping also results in the temporal focusing, limited only by the bandwidth of the control pulse. The influence of the exact value of $\epsilon $ appears to be rather insignificant, except for the periodic limit ($\epsilon=0.1\%$) where the penetration of the control pulse inside the sample is largely prohibited.

The calculated two-photon field intensities are shown in the lower plots of Fig. \ref{ff} for different arrival times of the pump pulse at the front layer, relative to the arrival time of the unshaped control pulse. For delay times earlier than around 2 ps, the two-photon intensity is not spatially localized, since the pump pulse overlaps largely with the forward traveling portion of the shaped control field. However, as soon as the pump is sufficiently delayed, its overlap with the backward propagating control results in a focused two-photon intensity (thin bright trace at the lower half of the plot). The lower limit of the spatial size of the focused two-photon intensity is again determined by the frequency bandwidth of the pump and control pulses. The focusing location can be tuned continuously across the front half of the sample. The target point of backscattering sets an upper limit for the position of the nonlinear focus. In the periodic limit, no control over the two-photon field can be achieved due to the inhibited light penetration into the stack.

Very similar one-photon control field trajectories and two-photon signal maps are plotted in Figures \ref{fb100} and \ref{fb0} using the two feedback-based shaping methods. In Fig. \ref{fb100}, we optimize the phase mask so as to maximize the intensity of the backward propagating control field at the location of the 50th air layer. In Fig. \ref{fb0}, the same adaptive optimization is then repeated with a more practical feedback parameter - the intensity of the backscattered control pulse as it exists the very front layer after a fixed delay of 6 ps, set in accordance with the results of the feedforward scheme. Note that the exact value of this delay is of little importance since the arrival time is easily controlled by adding a linear spectral phase tilt to the adaptive shaping mask. For both optimization methods, we used the ``minimize" algorithm from SciPy's ``optimize" library. The results demonstrate that both methods are capable of generating the backscattered pulse and producing the desired two-photon focusing similarly to the feedforward-based procedure. Furthermore, the methods proved equally successful for many different realizations of the stack geometry, indicating their robustness with respect to the randomness of the system.

\begin{figure*}[t]
\includegraphics[scale=0.5]{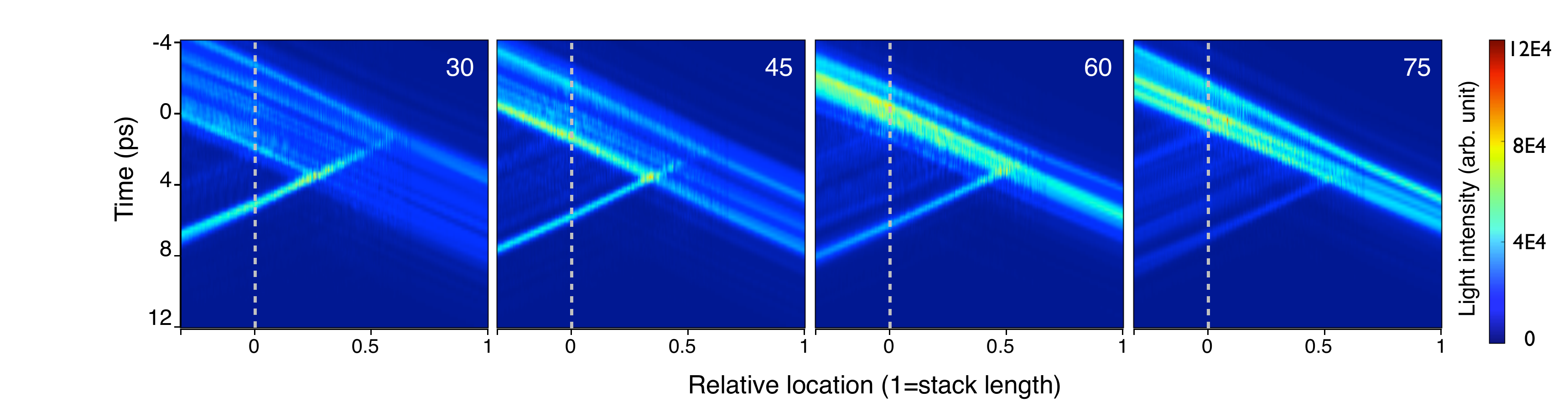}
\caption{(color online) The one-photon control-pulse intensity distribution for the \textit{feedforward} shaping method targeting 30th, 45th, 60th and 75th air layer.
}
\label{jtar}
\includegraphics[scale=0.46]{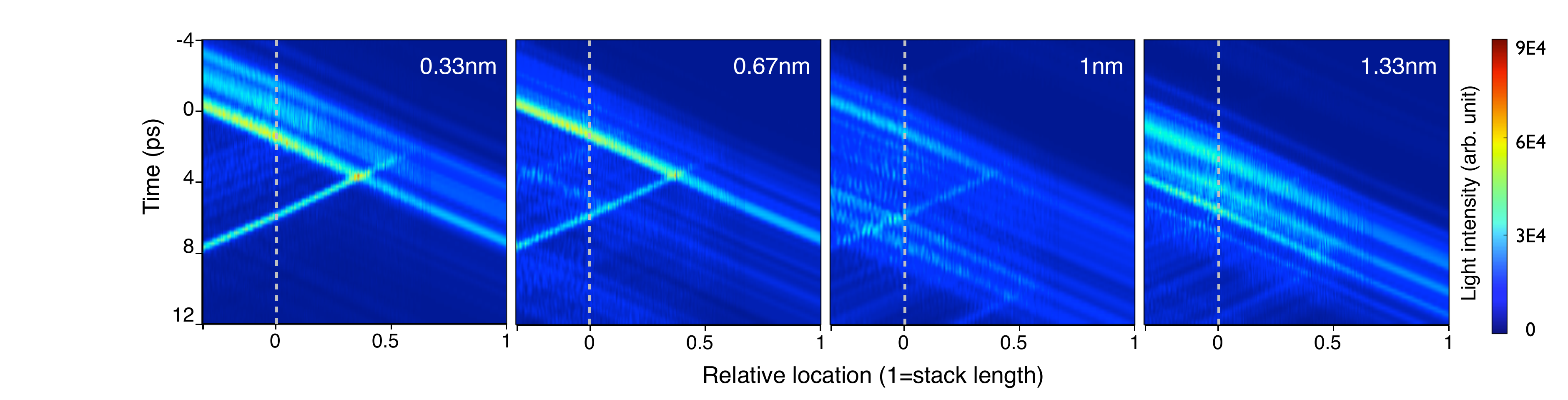}
\caption{(color online) The one-photon control-pulse intensity distribution for the \textit{feedforward} shaping method with worsening shaping resolution of 0.33~nm, 0.67~nm, 1~nm and 1.3~nm.
}
\label{res}
\end{figure*}

\subsubsection{Depth of Focusing and Shaping Resolution}

Since the enhanced backscattering, described above, is achieved via constructive interference of backward propagating waves at a target location, its efficiency should depend on the availability and strength of localized modes at that location. One would therefore expect the efficiency of our method to decrease when the reflection point of control pulses is pushed deeper into the stack (cf. Fig. \ref{mode_scan}\textbf{c}). This trend is apparent in Fig. \ref{jtar}
for a stack with $\epsilon=90\%$: as we move the target location from 30th to 45th, 60th and finally 75th air layer, the reflection point follows along, but the backscattered intensity is dropping. Hence, the control range for the focused two-photon field can be extended only at the expense of its lower strength.

Arranging for constructive interference within localized modes also depends on the spectral resolution, with which individual spectral phases can be varied. To analyze this dependence, we implement our feedforward method and change the resolution of a numerical pulse shaper from 0.07 nm to 0.33, 0.67, 1 and 1.3 nm, the latter being of the order of the average wavelength separation of the localized modes.
The results are shown in Fig. \ref{res} for a random stack with $\epsilon=20\%$. The detrimental effects of insufficient resolution to control individual spectral components within each single localized mode are clearly seen through the disappearance of the backscattered wave in the right most plot.

\section{Conclusion}

In conclusion, we have developed a systematic technique to achieve spatial focusing of a multi-photon field inside a 1D layered random medium. Our approach relies on creating a strong backscattered pulse originating from inside the random sample by means of the spectral pulse shaping. The method is demonstrated numerically for random stacks whose parameters are known \textit{a priori}, as well as using adaptive optimization algorithms that maximize the backward-propagating light intensity, either internally or at the very front layer of samples with unknown geometry. By colliding the controlled backscattered pulse with a time-delayed forward-propagating pump, spatially localized two-photon field is established in a range of target locations.

We have demonstrated the robustness of the proposed technique with respect to the degree of disorder. It works well both on a single-realization level and when statistically averaged over many realizations of the random medium. The location of the focused two-photon field can be easily controlled within (but not much beyond) the first half of the sample, as determined by an optimum reflection point for the spectrally shaped control pulse. This limit originates from the fundamental phenomenon of light localization in one-dimensional random structures. We also show that the required resolution of the spectral pulse shaping is dictated by the average frequency bandwidth of the localized modes. When the shaping resolution falls below this limit, the ability to control the interference of backward-propagating components is lost, and the efficiency of the proposed focusing method deteriorates.

One direct application, which served as a motivation for this study, is to induce and control local pumping inside a 1D random laser by means of the spectral pulse shaping and with no optical access other than the front layer of the sample. This ability should allow selective excitation of different lasing modes at different spatial locations, and would therefore provide control over the lasing frequencies. Other potential applications may include two-photon fluorescence imaging in layered structures, and can be extended to other scenarios of multi-photon light-matter interaction in random media.

\bibliography{han_milner}

\begin{thebibliography}{38}
\expandafter\ifx\csname natexlab\endcsname\relax\def\natexlab#1{#1}\fi
\expandafter\ifx\csname bibnamefont\endcsname\relax
  \def\bibnamefont#1{#1}\fi
\expandafter\ifx\csname bibfnamefont\endcsname\relax
  \def\bibfnamefont#1{#1}\fi
\expandafter\ifx\csname citenamefont\endcsname\relax
  \def\citenamefont#1{#1}\fi
\expandafter\ifx\csname url\endcsname\relax
  \def\url#1{\texttt{#1}}\fi
\expandafter\ifx\csname urlprefix\endcsname\relax\def\urlprefix{URL }\fi
\providecommand{\bibinfo}[2]{#2}
\providecommand{\eprint}[2][]{\url{#2}}

\bibitem[{\citenamefont{Letokhov}(1968)}]{Letokhov1968}
\bibinfo{author}{\bibfnamefont{V.~S.} \bibnamefont{Letokhov}},
  \bibinfo{journal}{Sov. Phys. JETP} \textbf{\bibinfo{volume}{26}},
  \bibinfo{pages}{835} (\bibinfo{year}{1968}).

\bibitem[{\citenamefont{Cao}(2003)}]{Cao2003}
\bibinfo{author}{\bibfnamefont{H.}~\bibnamefont{Cao}}, \bibinfo{journal}{Waves
  in random media} \textbf{\bibinfo{volume}{13}}, \bibinfo{pages}{R1}
  (\bibinfo{year}{2003}).

\bibitem[{\citenamefont{Wiersma}(2008)}]{Wiersma2008}
\bibinfo{author}{\bibfnamefont{D.~S.} \bibnamefont{Wiersma}},
  \bibinfo{journal}{Nat Phys} \textbf{\bibinfo{volume}{4}},
  \bibinfo{pages}{359} (\bibinfo{year}{2008}).

\bibitem[{\citenamefont{Genack and Drake}(1994)}]{Genack1994}
\bibinfo{author}{\bibfnamefont{A.~Z.} \bibnamefont{Genack}} \bibnamefont{and}
  \bibinfo{author}{\bibfnamefont{J.~M.} \bibnamefont{Drake}},
  \bibinfo{journal}{Nature} \textbf{\bibinfo{volume}{368}},
  \bibinfo{pages}{400} (\bibinfo{year}{1994}).

\bibitem[{\citenamefont{Lawandy et~al.}(1994)\citenamefont{Lawandy,
  Balachandran, Gomes, and Sauvain}}]{Lawandy1994}
\bibinfo{author}{\bibfnamefont{N.~M.} \bibnamefont{Lawandy}},
  \bibinfo{author}{\bibfnamefont{R.~M.} \bibnamefont{Balachandran}},
  \bibinfo{author}{\bibfnamefont{A.~S.~L.} \bibnamefont{Gomes}},
  \bibnamefont{and} \bibinfo{author}{\bibfnamefont{E.}~\bibnamefont{Sauvain}},
  \bibinfo{journal}{Nature} \textbf{\bibinfo{volume}{368}},
  \bibinfo{pages}{436} (\bibinfo{year}{1994}).

\bibitem[{\citenamefont{Sha et~al.}(1994)\citenamefont{Sha, Liu, and
  Alfano}}]{Sha1994}
\bibinfo{author}{\bibfnamefont{W.~L.} \bibnamefont{Sha}},
  \bibinfo{author}{\bibfnamefont{C.-H.} \bibnamefont{Liu}}, \bibnamefont{and}
  \bibinfo{author}{\bibfnamefont{R.~R.} \bibnamefont{Alfano}},
  \bibinfo{journal}{Opt. Lett.} \textbf{\bibinfo{volume}{19}},
  \bibinfo{pages}{1922} (\bibinfo{year}{1994}).

\bibitem[{\citenamefont{Noginov et~al.}(1995)\citenamefont{Noginov, Caulfield,
  Noginova, and Venkateswarlu}}]{Noginov1995}
\bibinfo{author}{\bibfnamefont{M.~A.} \bibnamefont{Noginov}},
  \bibinfo{author}{\bibfnamefont{H.~J.} \bibnamefont{Caulfield}},
  \bibinfo{author}{\bibfnamefont{N.~E.} \bibnamefont{Noginova}},
  \bibnamefont{and}
  \bibinfo{author}{\bibfnamefont{P.}~\bibnamefont{Venkateswarlu}},
  \bibinfo{journal}{Optics Communications} \textbf{\bibinfo{volume}{118}},
  \bibinfo{pages}{430} (\bibinfo{year}{1995}).

\bibitem[{\citenamefont{Wiersma et~al.}(1995)\citenamefont{Wiersma, Albada, and
  Lagendijk}}]{Wiersma1995}
\bibinfo{author}{\bibfnamefont{D.}~\bibnamefont{Wiersma}},
  \bibinfo{author}{\bibfnamefont{M.~P.~v.} \bibnamefont{Albada}},
  \bibnamefont{and}
  \bibinfo{author}{\bibfnamefont{A.}~\bibnamefont{Lagendijk}},
  \bibinfo{journal}{Nature} \textbf{\bibinfo{volume}{373}},
  \bibinfo{pages}{203} (\bibinfo{year}{1995}).

\bibitem[{\citenamefont{Cao et~al.}(1999)\citenamefont{Cao, Zhao, Ho, Seelig,
  Wang, and Chang}}]{Cao1999}
\bibinfo{author}{\bibfnamefont{H.}~\bibnamefont{Cao}},
  \bibinfo{author}{\bibfnamefont{Y.~G.} \bibnamefont{Zhao}},
  \bibinfo{author}{\bibfnamefont{S.~T.} \bibnamefont{Ho}},
  \bibinfo{author}{\bibfnamefont{E.~W.} \bibnamefont{Seelig}},
  \bibinfo{author}{\bibfnamefont{Q.~H.} \bibnamefont{Wang}}, \bibnamefont{and}
  \bibinfo{author}{\bibfnamefont{R.~P.~H.} \bibnamefont{Chang}},
  \bibinfo{journal}{PRL} \textbf{\bibinfo{volume}{82}}, \bibinfo{pages}{2278}
  (\bibinfo{year}{1999}).

\bibitem[{\citenamefont{Frolov et~al.}(1999)\citenamefont{Frolov, Vardeny,
  Yoshino, Zakhidov, and Baughman}}]{Frolov1999}
\bibinfo{author}{\bibfnamefont{S.~V.} \bibnamefont{Frolov}},
  \bibinfo{author}{\bibfnamefont{Z.~V.} \bibnamefont{Vardeny}},
  \bibinfo{author}{\bibfnamefont{K.}~\bibnamefont{Yoshino}},
  \bibinfo{author}{\bibfnamefont{A.}~\bibnamefont{Zakhidov}}, \bibnamefont{and}
  \bibinfo{author}{\bibfnamefont{R.~H.} \bibnamefont{Baughman}},
  \bibinfo{journal}{Phys. Rev. B} \textbf{\bibinfo{volume}{59}},
  \bibinfo{pages}{R5284} (\bibinfo{year}{1999}).

\bibitem[{\citenamefont{Jiang and Soukoulis}(2000)}]{Jiang2000}
\bibinfo{author}{\bibfnamefont{X.}~\bibnamefont{Jiang}} \bibnamefont{and}
  \bibinfo{author}{\bibfnamefont{C.~M.} \bibnamefont{Soukoulis}},
  \bibinfo{journal}{Physical Review Letters} \textbf{\bibinfo{volume}{85}},
  \bibinfo{pages}{70} (\bibinfo{year}{2000}).

\bibitem[{\citenamefont{Andreasen et~al.}(2011)\citenamefont{Andreasen,
  Asatryan, Botten, Byrne, Cao, Ge, Labonté, Sebbah, Stone, Türeci
  et~al.}}]{Andreasen2011}
\bibinfo{author}{\bibfnamefont{J.}~\bibnamefont{Andreasen}},
  \bibinfo{author}{\bibfnamefont{A.~A.} \bibnamefont{Asatryan}},
  \bibinfo{author}{\bibfnamefont{L.~C.} \bibnamefont{Botten}},
  \bibinfo{author}{\bibfnamefont{M.~A.} \bibnamefont{Byrne}},
  \bibinfo{author}{\bibfnamefont{H.}~\bibnamefont{Cao}},
  \bibinfo{author}{\bibfnamefont{L.}~\bibnamefont{Ge}},
  \bibinfo{author}{\bibfnamefont{L.}~\bibnamefont{Labonté}},
  \bibinfo{author}{\bibfnamefont{P.}~\bibnamefont{Sebbah}},
  \bibinfo{author}{\bibfnamefont{A.~D.} \bibnamefont{Stone}},
  \bibinfo{author}{\bibfnamefont{H.~E.} \bibnamefont{Türeci}},
  \bibnamefont{et~al.}, \bibinfo{journal}{Adv. Opt. Photon.}
  \textbf{\bibinfo{volume}{3}}, \bibinfo{pages}{88} (\bibinfo{year}{2011}).

\bibitem[{\citenamefont{Wiersma and Cavalieri}(2001)}]{Wiersma2001}
\bibinfo{author}{\bibfnamefont{D.~S.} \bibnamefont{Wiersma}} \bibnamefont{and}
  \bibinfo{author}{\bibfnamefont{S.}~\bibnamefont{Cavalieri}},
  \bibinfo{journal}{Nature} \textbf{\bibinfo{volume}{414}},
  \bibinfo{pages}{708} (\bibinfo{year}{2001}).

\bibitem[{\citenamefont{Gottardo et~al.}(2008)\citenamefont{Gottardo, Sapienza,
  Garcia, Blanco, Wiersma, and Lopez}}]{Gottardo2008}
\bibinfo{author}{\bibfnamefont{S.}~\bibnamefont{Gottardo}},
  \bibinfo{author}{\bibfnamefont{R.}~\bibnamefont{Sapienza}},
  \bibinfo{author}{\bibfnamefont{P.~D.} \bibnamefont{Garcia}},
  \bibinfo{author}{\bibfnamefont{A.}~\bibnamefont{Blanco}},
  \bibinfo{author}{\bibfnamefont{D.~S.} \bibnamefont{Wiersma}},
  \bibnamefont{and} \bibinfo{author}{\bibfnamefont{C.}~\bibnamefont{Lopez}},
  \bibinfo{journal}{Nat Photon} \textbf{\bibinfo{volume}{2}},
  \bibinfo{pages}{429} (\bibinfo{year}{2008}).

\bibitem[{\citenamefont{Bardoux et~al.}(2011)\citenamefont{Bardoux, Kaneta,
  Funato, Okamoto, Kawakami, Kikuchi, and Kishino}}]{Bardoux2011}
\bibinfo{author}{\bibfnamefont{R.}~\bibnamefont{Bardoux}},
  \bibinfo{author}{\bibfnamefont{A.}~\bibnamefont{Kaneta}},
  \bibinfo{author}{\bibfnamefont{M.}~\bibnamefont{Funato}},
  \bibinfo{author}{\bibfnamefont{K.}~\bibnamefont{Okamoto}},
  \bibinfo{author}{\bibfnamefont{Y.}~\bibnamefont{Kawakami}},
  \bibinfo{author}{\bibfnamefont{A.}~\bibnamefont{Kikuchi}}, \bibnamefont{and}
  \bibinfo{author}{\bibfnamefont{K.}~\bibnamefont{Kishino}},
  \bibinfo{journal}{Opt. Express} \textbf{\bibinfo{volume}{19}},
  \bibinfo{pages}{9262} (\bibinfo{year}{2011}).

\bibitem[{\citenamefont{El-Dardiry and Lagendijk}(2011)}]{ElDardiry2011}
\bibinfo{author}{\bibfnamefont{R.~G.~S.} \bibnamefont{El-Dardiry}}
  \bibnamefont{and}
  \bibinfo{author}{\bibfnamefont{A.}~\bibnamefont{Lagendijk}},
  \bibinfo{journal}{Applied Physics Letters} \textbf{\bibinfo{volume}{98}},
  \bibinfo{pages}{161106} (\bibinfo{year}{2011}).

\bibitem[{\citenamefont{Shojaie et~al.}(2014)\citenamefont{Shojaie, Mirzaei,
  and Bahrampour}}]{Shojaie2014}
\bibinfo{author}{\bibfnamefont{E.}~\bibnamefont{Shojaie}},
  \bibinfo{author}{\bibfnamefont{A.}~\bibnamefont{Mirzaei}}, \bibnamefont{and}
  \bibinfo{author}{\bibfnamefont{A.}~\bibnamefont{Bahrampour}},
  \bibinfo{journal}{Optics Letters} \textbf{\bibinfo{volume}{39}},
  \bibinfo{pages}{4537} (\bibinfo{year}{2014}).

\bibitem[{\citenamefont{Wu et~al.}(2007)\citenamefont{Wu, Andreasen, Cao, and
  Yamilov}}]{Wu2007}
\bibinfo{author}{\bibfnamefont{X.}~\bibnamefont{Wu}},
  \bibinfo{author}{\bibfnamefont{J.}~\bibnamefont{Andreasen}},
  \bibinfo{author}{\bibfnamefont{H.}~\bibnamefont{Cao}}, \bibnamefont{and}
  \bibinfo{author}{\bibfnamefont{A.}~\bibnamefont{Yamilov}},
  \bibinfo{journal}{J. Opt. Soc. Am. B} \textbf{\bibinfo{volume}{24}},
  \bibinfo{pages}{A26} (\bibinfo{year}{2007}).

\bibitem[{\citenamefont{Bachelard et~al.}(2012)\citenamefont{Bachelard,
  Andreasen, Gigan, and Sebbah}}]{Bachelard2012}
\bibinfo{author}{\bibfnamefont{N.}~\bibnamefont{Bachelard}},
  \bibinfo{author}{\bibfnamefont{J.}~\bibnamefont{Andreasen}},
  \bibinfo{author}{\bibfnamefont{S.}~\bibnamefont{Gigan}}, \bibnamefont{and}
  \bibinfo{author}{\bibfnamefont{P.}~\bibnamefont{Sebbah}},
  \bibinfo{journal}{Physical Review Letters} \textbf{\bibinfo{volume}{109}},
  \bibinfo{pages}{033903} (\bibinfo{year}{2012}).

\bibitem[{\citenamefont{Hisch et~al.}(2013)\citenamefont{Hisch, Liertzer,
  Pogany, Mintert, and Rotter}}]{Hisch2013}
\bibinfo{author}{\bibfnamefont{T.}~\bibnamefont{Hisch}},
  \bibinfo{author}{\bibfnamefont{M.}~\bibnamefont{Liertzer}},
  \bibinfo{author}{\bibfnamefont{D.}~\bibnamefont{Pogany}},
  \bibinfo{author}{\bibfnamefont{F.}~\bibnamefont{Mintert}}, \bibnamefont{and}
  \bibinfo{author}{\bibfnamefont{S.}~\bibnamefont{Rotter}},
  \bibinfo{journal}{Physical Review Letters} \textbf{\bibinfo{volume}{111}},
  \bibinfo{pages}{023902} (\bibinfo{year}{2013}).

\bibitem[{\citenamefont{Vanneste and Sebbah}(2001)}]{Vanneste2001}
\bibinfo{author}{\bibfnamefont{C.}~\bibnamefont{Vanneste}} \bibnamefont{and}
  \bibinfo{author}{\bibfnamefont{P.}~\bibnamefont{Sebbah}},
  \bibinfo{journal}{Physical Review Letters} \textbf{\bibinfo{volume}{87}},
  \bibinfo{pages}{183903} (\bibinfo{year}{2001}).

\bibitem[{\citenamefont{Sebbah and Vanneste}(2002)}]{Sebbah2002}
\bibinfo{author}{\bibfnamefont{P.}~\bibnamefont{Sebbah}} \bibnamefont{and}
  \bibinfo{author}{\bibfnamefont{C.}~\bibnamefont{Vanneste}},
  \bibinfo{journal}{Physical Review B} \textbf{\bibinfo{volume}{66}},
  \bibinfo{pages}{144202} (\bibinfo{year}{2002}).

\bibitem[{\citenamefont{Burin et~al.}(2002)\citenamefont{Burin, Ratner, Cao,
  and Chang}}]{Burin2002}
\bibinfo{author}{\bibfnamefont{A.~L.} \bibnamefont{Burin}},
  \bibinfo{author}{\bibfnamefont{M.~A.} \bibnamefont{Ratner}},
  \bibinfo{author}{\bibfnamefont{H.}~\bibnamefont{Cao}}, \bibnamefont{and}
  \bibinfo{author}{\bibfnamefont{S.~H.} \bibnamefont{Chang}},
  \bibinfo{journal}{Physical Review Letters} \textbf{\bibinfo{volume}{88}},
  \bibinfo{pages}{093904} (\bibinfo{year}{2002}).

\bibitem[{\citenamefont{Milner and Genack}(2005)}]{Milner2005}
\bibinfo{author}{\bibfnamefont{V.}~\bibnamefont{Milner}} \bibnamefont{and}
  \bibinfo{author}{\bibfnamefont{A.~Z.} \bibnamefont{Genack}},
  \bibinfo{journal}{Physical Review Letters} \textbf{\bibinfo{volume}{94}},
  \bibinfo{pages}{073901} (\bibinfo{year}{2005}).

\bibitem[{\citenamefont{Bachelard et~al.}(2014)\citenamefont{Bachelard, Gigan,
  Noblin, and Sebbah}}]{Bachelard2014}
\bibinfo{author}{\bibfnamefont{N.}~\bibnamefont{Bachelard}},
  \bibinfo{author}{\bibfnamefont{S.}~\bibnamefont{Gigan}},
  \bibinfo{author}{\bibfnamefont{X.}~\bibnamefont{Noblin}}, \bibnamefont{and}
  \bibinfo{author}{\bibfnamefont{P.}~\bibnamefont{Sebbah}},
  \bibinfo{journal}{Nat Phys} \textbf{\bibinfo{volume}{10}},
  \bibinfo{pages}{426} (\bibinfo{year}{2014}).

\bibitem[{\citenamefont{Aulbach et~al.}(2011)\citenamefont{Aulbach, Gjonaj,
  Johnson, Mosk, and Lagendijk}}]{Aulbach2011}
\bibinfo{author}{\bibfnamefont{J.}~\bibnamefont{Aulbach}},
  \bibinfo{author}{\bibfnamefont{B.}~\bibnamefont{Gjonaj}},
  \bibinfo{author}{\bibfnamefont{P.~M.} \bibnamefont{Johnson}},
  \bibinfo{author}{\bibfnamefont{A.~P.} \bibnamefont{Mosk}}, \bibnamefont{and}
  \bibinfo{author}{\bibfnamefont{A.}~\bibnamefont{Lagendijk}},
  \bibinfo{journal}{Physical Review Letters} \textbf{\bibinfo{volume}{106}},
  \bibinfo{pages}{103901} (\bibinfo{year}{2011}).

\bibitem[{\citenamefont{Katz et~al.}(2011)\citenamefont{Katz, Small, Bromberg,
  and Silberberg}}]{Katz2011}
\bibinfo{author}{\bibfnamefont{O.}~\bibnamefont{Katz}},
  \bibinfo{author}{\bibfnamefont{E.}~\bibnamefont{Small}},
  \bibinfo{author}{\bibfnamefont{Y.}~\bibnamefont{Bromberg}}, \bibnamefont{and}
  \bibinfo{author}{\bibfnamefont{Y.}~\bibnamefont{Silberberg}},
  \bibinfo{journal}{Nat Photon} \textbf{\bibinfo{volume}{5}},
  \bibinfo{pages}{372} (\bibinfo{year}{2011}).

\bibitem[{\citenamefont{Mosk et~al.}(2012)\citenamefont{Mosk, Lagendijk,
  Lerosey, and Fink}}]{Mosk2012}
\bibinfo{author}{\bibfnamefont{A.~P.} \bibnamefont{Mosk}},
  \bibinfo{author}{\bibfnamefont{A.}~\bibnamefont{Lagendijk}},
  \bibinfo{author}{\bibfnamefont{G.}~\bibnamefont{Lerosey}}, \bibnamefont{and}
  \bibinfo{author}{\bibfnamefont{M.}~\bibnamefont{Fink}}, \bibinfo{journal}{Nat
  Photon} \textbf{\bibinfo{volume}{6}}, \bibinfo{pages}{283}
  (\bibinfo{year}{2012}).

\bibitem[{\citenamefont{van Beijnum et~al.}(2011)\citenamefont{van Beijnum, van
  Putten, Lagendijk, and Mosk}}]{VanBeijnum2011}
\bibinfo{author}{\bibfnamefont{F.}~\bibnamefont{van Beijnum}},
  \bibinfo{author}{\bibfnamefont{E.~G.} \bibnamefont{van Putten}},
  \bibinfo{author}{\bibfnamefont{A.}~\bibnamefont{Lagendijk}},
  \bibnamefont{and} \bibinfo{author}{\bibfnamefont{A.~P.} \bibnamefont{Mosk}},
  \bibinfo{journal}{Optics Letters} \textbf{\bibinfo{volume}{36}},
  \bibinfo{pages}{373} (\bibinfo{year}{2011}).

\bibitem[{\citenamefont{McCabe et~al.}(2011)\citenamefont{McCabe, Tajalli,
  Austin, Bondareff, Walmsley, Gigan, and Chatel}}]{McCabe2011}
\bibinfo{author}{\bibfnamefont{D.~J.} \bibnamefont{McCabe}},
  \bibinfo{author}{\bibfnamefont{A.}~\bibnamefont{Tajalli}},
  \bibinfo{author}{\bibfnamefont{D.~R.} \bibnamefont{Austin}},
  \bibinfo{author}{\bibfnamefont{P.}~\bibnamefont{Bondareff}},
  \bibinfo{author}{\bibfnamefont{I.~A.} \bibnamefont{Walmsley}},
  \bibinfo{author}{\bibfnamefont{S.}~\bibnamefont{Gigan}}, \bibnamefont{and}
  \bibinfo{author}{\bibfnamefont{B.}~\bibnamefont{Chatel}},
  \bibinfo{journal}{Nat Commun} \textbf{\bibinfo{volume}{2}},
  \bibinfo{pages}{447} (\bibinfo{year}{2011}).

\bibitem[{\citenamefont{Meshulach and Silberberg}(1998)}]{Meshulach1998}
\bibinfo{author}{\bibfnamefont{D.}~\bibnamefont{Meshulach}} \bibnamefont{and}
  \bibinfo{author}{\bibfnamefont{Y.}~\bibnamefont{Silberberg}},
  \bibinfo{journal}{Nature} \textbf{\bibinfo{volume}{396}},
  \bibinfo{pages}{239} (\bibinfo{year}{1998}).

\bibitem[{\citenamefont{Dela~Cruz et~al.}(2004)\citenamefont{Dela~Cruz,
  Pastirk, Comstock, Lozovoy, and Dantus}}]{Delacruz2004}
\bibinfo{author}{\bibfnamefont{J.~M.} \bibnamefont{Dela~Cruz}},
  \bibinfo{author}{\bibfnamefont{I.}~\bibnamefont{Pastirk}},
  \bibinfo{author}{\bibfnamefont{M.}~\bibnamefont{Comstock}},
  \bibinfo{author}{\bibfnamefont{V.~V.} \bibnamefont{Lozovoy}},
  \bibnamefont{and} \bibinfo{author}{\bibfnamefont{M.}~\bibnamefont{Dantus}},
  \bibinfo{journal}{Proceedings of the National Academy of Sciences of the
  United States of America} \textbf{\bibinfo{volume}{101}},
  \bibinfo{pages}{16996} (\bibinfo{year}{2004}).

\bibitem[{\citenamefont{Berry and Klein}(1997)}]{Berry1997}
\bibinfo{author}{\bibfnamefont{M.~V.} \bibnamefont{Berry}} \bibnamefont{and}
  \bibinfo{author}{\bibfnamefont{S.}~\bibnamefont{Klein}},
  \bibinfo{journal}{European Journal of Physics} \textbf{\bibinfo{volume}{18}},
  \bibinfo{pages}{222} (\bibinfo{year}{1997}).

\bibitem[{\citenamefont{Weiner}(2000)}]{Weiner2000}
\bibinfo{author}{\bibfnamefont{A.~M.} \bibnamefont{Weiner}},
  \bibinfo{journal}{Review of Scientific Instruments}
  \textbf{\bibinfo{volume}{71}}, \bibinfo{pages}{1929} (\bibinfo{year}{2000}).

\bibitem[{\citenamefont{Drane et~al.}(2015)\citenamefont{Drane, Bach, Shapiro,
  and Milner}}]{Drane2015}
\bibinfo{author}{\bibfnamefont{T.~M.} \bibnamefont{Drane}},
  \bibinfo{author}{\bibfnamefont{H.}~\bibnamefont{Bach}},
  \bibinfo{author}{\bibfnamefont{M.}~\bibnamefont{Shapiro}}, \bibnamefont{and}
  \bibinfo{author}{\bibfnamefont{V.}~\bibnamefont{Milner}},
  \bibinfo{journal}{Biomedical Optics Express} \textbf{\bibinfo{volume}{6}},
  \bibinfo{pages}{1885} (\bibinfo{year}{2015}).

\bibitem[{\citenamefont{Meshulach and Silberberg}(1999)}]{Meshulach1999}
\bibinfo{author}{\bibfnamefont{D.}~\bibnamefont{Meshulach}} \bibnamefont{and}
  \bibinfo{author}{\bibfnamefont{Y.}~\bibnamefont{Silberberg}},
  \bibinfo{journal}{Physical Review A (Atomic, Molecular, and Optical Physics)}
  \textbf{\bibinfo{volume}{60}}, \bibinfo{pages}{1287} (\bibinfo{year}{1999}).

\bibitem[{\citenamefont{van~der Ziel et~al.}(1981)\citenamefont{van~der Ziel,
  Logan, and Mikulyak}}]{vanderZiel1981}
\bibinfo{author}{\bibfnamefont{J.~P.} \bibnamefont{van~der Ziel}},
  \bibinfo{author}{\bibfnamefont{R.~A.} \bibnamefont{Logan}}, \bibnamefont{and}
  \bibinfo{author}{\bibfnamefont{R.~M.} \bibnamefont{Mikulyak}},
  \bibinfo{journal}{Applied Physics Letters} \textbf{\bibinfo{volume}{39}},
  \bibinfo{pages}{867} (\bibinfo{year}{1981}).

\bibitem[{\citenamefont{Pendry}(1994)}]{Pendry1994}
\bibinfo{author}{\bibfnamefont{J.~B.} \bibnamefont{Pendry}},
  \bibinfo{journal}{Advances in Physics} \textbf{\bibinfo{volume}{43}},
  \bibinfo{pages}{461} (\bibinfo{year}{1994}).

\end{thebibliography}

\end{document}